\documentclass[eqsecnum,showpacs,aps,epsf,twocolumn,nofootinbib]{revtex4}
\usepackage{graphicx}
\usepackage{amsmath}

\begin{document}



\title{Creation of a brane world with a bulk scalar field}

\author{
Koh-suke Aoyanagi$^1$\footnote{E-mail
address : aoyanagi@gravity.phys.waseda.ac.jp}}
\author{
Kei-ichi Maeda$^{1,2,3}$\footnote{E-mail address : maeda@gravity.phys.waseda.ac.jp}
}

\address{$^{1}$Department of Physics, Waseda University,
Shinjuku-ku, Tokyo 169-8555, Japan~}
\address{$^2$Advanced Research Institute for Science and Engineering,
Waseda  University, Shinjuku-ku, Tokyo 169-8555, Japan~}
\address{$^3$Waseda Institute for Astrophysics, Waseda University,
Shinjuku-ku,  Tokyo 169-8555, Japan~}

\date{\today}

\begin{abstract}
We investigate the creation of a brane world with a bulk scalar field.
We consider an exponential potential of a bulk scalar field: 
$V(\phi)\propto \exp(-2\beta \phi)$, 
where $\beta$ is the parameter of the theory.
This model is based on a supersymmetric theory, and 
includes the Randall-Sundrum model ($\beta=0$) and the
5-dimensional effective model of the 
Ho\u{r}va-Witten theory ($\beta=1$).
We show that for this potential a brane instanton is constructed 
only when the curvature of a brane vanishes, that is, the brane is flat.
We construct an instanton with two branes 
and a singular instanton with a single brane.
The Euclidean action of the singular instanton solution 
is finite if $\beta^2 >2/3$.
We also calculate perturbations of the action around a singular instanton 
solution in order to 
show that the singular instanton is well-defined.
\end{abstract}



\maketitle


\section{INTRODUCTION}
A new paradigm of cosmology based on
superstring/M-theory, the so-called ``brane world'',
has been discussed for last several years.
The prototype of a brane world was first discussed in \cite{Akama,
Rubakov-Shaposhnikov}.
More recently, this prototype has been combined with the idea of the
D-brane found by Polchinski in string 
theory \cite{Polchinski}, and a new paradigm of a brane world has been
developed \cite{Arkani,Randall-Sundrum99_1,Randall-Sundrum99_2}.
Here, one of the most interesting approaches is that of
Randall and Sundrum \cite{Randall-Sundrum99_1,Randall-Sundrum99_2}.
They considered a  pure 5-dimensional (5D) Einstein gravity 
only with a cosmological constant in a bulk.
In this scenario the effective 4D Einstein equations are obtained
by projecting the 5D metric onto the brane
\cite{Binetruy-Deffayet-Langlois00,Shiromizu-Maeda-Sasaki00}.

Whereas in a string/M-theory context
one would also expect additional scalar fields,
associated with the many moduli fields, as fundamental fields. 
Those fields will, in principle, 
also propagate in the bulk.
For example, Lukas, Ovrut and Waldram \cite{Lukas-Ovrut-Waldram99}
 derived an effective 5D action by a dimensional reduction
from 11-dimensional M-theory.
It  contains a scalar field in the 5D bulk,
which corresponds to a moduli associated with compactification
of six extra dimensions onto a Calabi-Yau space.
In the model of a brane world with a bulk scalar fields, 
the effective 4-dimensional Einstein equations are
obtained covariantly by Maeda and Wands \cite{Maeda-Wands00}.
A brane world inflation with bulk scalar fields 
is discussed by Himemoto and his collaborators \cite{Himemoto-etal}.
They proposed a ``bulk inflaton model'' in which the inflation on the brane is
caused by a scalar field in the bulk, but the bulk itself is not inflating.
Cosmological perturbations are also discussed based on
dilatonic brane worlds \cite{Koyama-etal}.
The bulk scalar
field yields a power-law inflation on the brane.
In this model, perturbation equations are solved analytically.
Another analysis of cosmological perturbations in the bulk inflaton
models with an exponential potential is given in \cite{Kobayashi-Tanaka04}. 

If a brane world  describes our universe,
we need to consider the creation of a brane world.
Up to now many authors have studied the creation 
of the Universe in 4-dimensional spacetime.
First the quantum creation of the universe was 
suggested by Vilenkin \cite{Vilenkin82}.
This approach is based on the picture
that the Universe spontaneously nucleates
in a de Sitter space.
The mathematical description of this nucleation
is analogous to quantum tunneling 
through a potential barrier \cite{Coleman-deLuccia80}.
Another approach to quantum cosmology was developed by
Hartle and Hawking \cite{Hartle-Hawking83}.
They proposed that the wave function
of the Universe is given by a path integral 
over compact Euclidean geometries
with an appropriate boundary condition,
which is called a ``no boundary'' boundary condition.
Here the wave function of the Universe is 
expected to be proportional to $e^{-S_E}$,
where $S_E$ is the Euclidean action.

In this paper
we consider the creation of a brane world with a bulk scalar field 
using an instanton solution which is given by
solving the 5D Euclidean Einstein equations.
In order to construct a compact Euclidean manifold,
we have to  glue two copies of  a finite patch of a bulk 
spacetime with a brane boundary by use of Israel's junction
condition \cite{Israel66}.
Garriga and Sasaki  \cite{Garriga-Sasaki00} first constructed an
instanton including an inflating brane
in the Randall-Sundrum model.
Hawking, Hertog, and Reall
consider the creation of a brane world
using an instanton,
and discuss  inflation and 
fluctuations during the de Sitter phase in the model 
containing the quantum correction term called a trace anomaly on the brane
\cite{Hawking-Hetrog-Reall01,Hawking-Hetrog-Reall00}.
A model with similar quantum
corrections was also analyzed by Nojiri and Odintsov
\cite{Nojiri-Odintsov-Zerbini00}, 
and instanton solutions with
a Gauss-Bonnet term in the bulk were discussed in
\cite{Aoyanagi-Maeda04,Nojiri-Odintsov00}.

The plan of this paper is as follows.
In Sec. \ref{action-eqs},
we present the Euclidean action and 
Euclidean equations of motion in 
the brane model with a bulk scalar field.
In Sec. \ref{flat-brane},
we obtain an instanton solution 
with two branes and that with a single brane.
We then evaluate the Euclidean action 
of these instanton solutions.
Although the single-brane instanton has 
a singularity, we find that  the action is finite.
In Sec. \ref{2nd-action},
in order to show that this singularity is harmless,
we consider  perturbations of the  action around the instanton
solution.
Our conclusions and remarks follow in Sec. \ref{conclusion}.

\section{ACTION AND EQUATIONS OF MOTION} \label{action-eqs}
\begin{figure}[t]
\begin{center}
\begin{tabular}{c}
 \includegraphics[height=4cm]{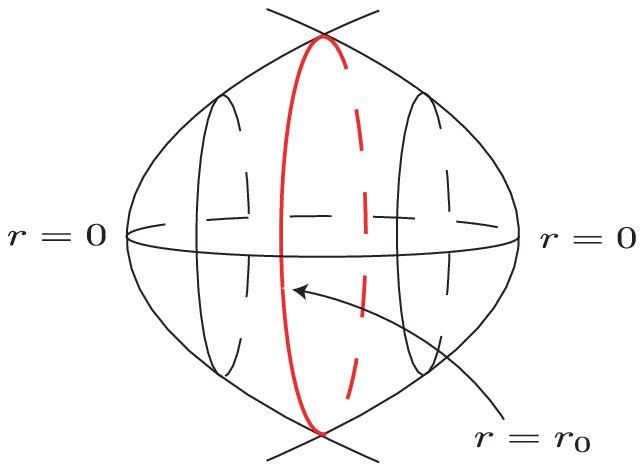} \\ (a) \\
 \includegraphics[height=4cm]{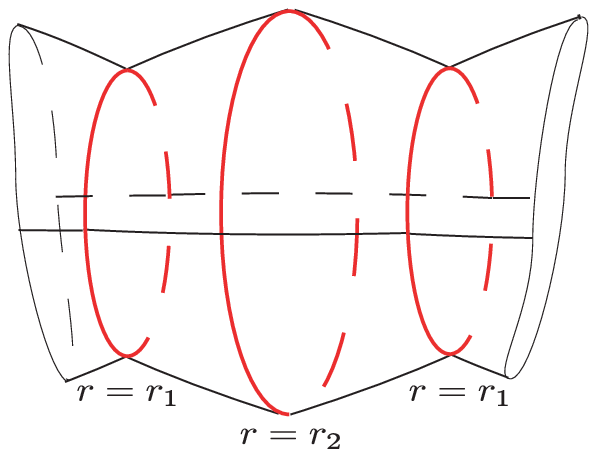} \\
 (b)\\
\end{tabular}
 \caption[fig1]{Schematic diagrams of brane instantons. 
We construct instanton solutions by gluing two copies of  a finite patch
of a bulk spacetime with a brane boundary by use of the  junction
condition. The thick vertical
circle at $r=r_i (i=0,1,2)$ represents the 4-dimensinal
maximally 
symmetric brane at which
the two identical 5-dimensional bulk spaces are glued.} \label{fig1}
\end{center}
\end{figure}

We consider  a scalar field in a bulk as well as gravity.
The Euclidean action is given by
\begin{eqnarray}
 S_E = - \frac{1}{2 \kappa^2_5} \left\{ \int_{{\mathcal M}} d^{5}x \sqrt{g} 
\left[ 
 R - \frac{1}{2} \partial_a \phi \partial^a \phi - V(\phi)
 \right] \right.  \nonumber  \\
 \left.
 - \sum_{i}\int_{r=r_i}d^4x \sqrt{g_{(i)}} 
 \left[ K^{\pm}_{(i)} - \lambda_{(i)}(\phi)\right] \right\}, \label{2-k}
\end{eqnarray}
where $\kappa^2_5$ is the 5-dimensional gravitational constant,
$r_i$ is the position of 
the $i$-th 3-brane,
$\lambda_{(i)}$ denotes  its tension,  and $K^{\pm}_{(i)}$
denotes its extrinsic curvature. 
The suffix $i$ corresponds to the numbering of branes. 
For a two-brane model, $i=1$ and $i=2$ stand for
a negative and a positive  tension brane, respectively.
Also for a single brane model, 
we use $i=0$ to stand for a brane (see Fig. \ref{fig1}).
  
Since we are looking for an instanton solution, we assume a highly
symmetric Euclidean spacetime, whose metric is given by
\begin{eqnarray}
 ds^2_E = dr^2+b(r)^2 \gamma_{\mu \nu} dx^{\mu}dx^{\nu}, \label{2-l}
\end{eqnarray}
where $\gamma_{\mu \nu}$ is the metric of 4-dimensional maximally
symmetric pace, which  is classified into three
types by the signature of curvature, i.e., $k=$0 (zero), 1 (positive), or
$-1$ (negative). These correspond to the curvature sign of the
Friedmann universe after creation.
Since the Euclidean space must be compact when we discuss its creation,
in the case of $k$=0 or $-1$, we have to make a space compact by identification.
Then the flat  spacetime is a  4-dimensional torus, and that with
 $k=-1$ has a more complicated topology.

The equations of motion with these ansatz are given by
\begin{eqnarray}
&&(b')^2 = k+ \frac{1}{12}b^2 \left(\frac{1}{2}(\phi')^2-V(\phi)
 \right),
\\
&&b'' =-\frac{1}{12}b\left[\frac{3}{2}(\phi')^2+V(\phi) 
 +2\lambda_{(i)}(\phi) \delta(r-r_i) \right],
\\
&& \phi'' +4\frac{b'}{b}\phi'=\frac{dV}{d\phi} 
  +\frac{d\lambda_{(i)}(\phi)}{d\phi} \delta(r-r_i),
\end{eqnarray}
where the prime denotes the derivative with respect to $r$, 
and $k$ is a curvature of the brane.

In what follows, we specify the form of 
a scalar field potential.
We assume that
the bulk potential for a scalar field 
and the tension of the $i$-th brane are given by
\begin{eqnarray}
 V(\phi) &=& \left(\beta^2 -\frac{2}{3} \right) v^2 e^{-2\beta
 \phi}, \label{2-h}
\\
  \lambda_i (\phi)&=& \pm 2\sqrt{2} v e^{-\beta \phi}, \label{2-j}
\end{eqnarray}
where $v (>0)$ describes a typical energy scale, 
and $\beta$ is a parameter of the model
\cite{Cvetic-Lu-Pope01, kobayashi-koyama02}.
The signature of Eq. (\ref{2-j}) corresponds to
 the sign of tension of a brane.

This form includes the Randall-Sundrum model ($\beta=0$) 
and the 5-dimensional effective model
derived from
M-theory via the Ho\u{r}ava-Witten theory
by Lukas, Ovrut and Waldram ($\beta =1$) \cite{Lukas-Ovrut-Waldram99}.
If we assume supersymmetry and the super potential
given by an exponential potential, we find the above forms of 
a scalar field potential in a bulk and a tension on a brane.
Hence, our ansatz is rather generic in the context of
a supersymmetric theory.
This form of potential has also been used in 
\cite{Cvetic-Lu-Pope01, kobayashi-koyama02}.

By using the above  potential and tension,
we rewrite our basic equations.
The equations of motion in the bulk are
\begin{eqnarray}
 &&
(b')^2 = k+ \frac{1}{12}b^2 \left[\frac{1}{2} (\phi')^2
 - \left(\beta^2 - \frac{2}{3} \right)
 v^2 e^{-2\beta \phi} \right], \label{2-a}
\\
&&
 b''=-\frac{1}{12}b\left[\frac{3}{2}(\phi')^2+  \left(\beta^2
 - \frac{2}{3} \right)v^2 e^{-2\beta \phi} \right], \label{2-b}
\\
&&
 \phi''+4\frac{b'}{b}\phi'=-2\beta  \left(\beta^2 
 - \frac{2}{3} \right)v^2 e^{-2\beta \phi}. \label{2-c}
\end{eqnarray}
Eqs. (\ref{2-b}) and (\ref{2-c}) are dynamical equations, 
while Eq. (\ref{2-a}) is a constraint equation.

On each brane, we have to impose a boundary condition.
Here we assume $Z_2$ symmetry on each brane as
a conventional brane world model.
If we adopt a different condition, our result may be changed.

From $Z_2$ symmetry, we have jump conditions for
$ b'$ and $ \phi'$ on each brane as
\begin{eqnarray}
 b'(r_i) &=& \mp \epsilon_i \frac{\sqrt{2}}{6}b(r_i) v e^{-\phi}
  \Bigm|_{r=r_{i}}, \label{2-d}
\\
  \phi'(r_i) &=& \mp \epsilon_i \sqrt{2}v e^{-\phi}  \Bigm|_{r=r_{i}}\,,
\label{2-e}
\end{eqnarray}
where the upper (lower) sign applies to the first boundary at $r=r_1$
(the second boundary at $r=r_2$)
and $\epsilon_i=\pm 1$ corresponds to the sign of tension
of the $i$-th brane. 
For a single-brane model, we take the lower sign. 

For a single-brane model, 
we also 
have to impose another boundary
condition at $r=0$. Here we adopt
the ``no boundary boundary condition''
\cite{Hartle-Hawking83}.
This condition comes from regularity of the 5-dimensional
geometry. 
For $k=1$, it   
simply gives 
the boundary condition of 
$b(0)=0$, $b'(0)=1$, $\phi'(0)=0$.
However,  for $k=0$ and $k=-1$,
the same boundary condition
does not give a regular spacetime.
Hence we may have to impose a 
different boundary condition such as
$b(0)\neq 0$, $b'(0)=0$ with identification at $r=0$.
That is,
we have to consider 
5-dimensional compactified manifold with 
different topology
from that in Fig.~\ref{fig1}(a)
such as a 5-dimensional inhomogeneous torus
with a brane at $r=r_0$. 
In this paper, we just focus on instanton solutions 
with topology shown in Fig.~\ref{fig1}(a).
For instanton solutions with a different topology,
we will analyze in a separated paper.

Although the condition $b(0)=0$ for $k=0$ and $k=-1$
causes a singularity as mentioned above, 
we may be able to construct a singular instanton such as
the Hawking-Turok instanton \cite{Hawking-Turok98}.
Then we also discuss a singular instanton in this paper. 

Before constructing an instanton solution,
 we show that a brane must be flat.
The constraint equation (\ref{2-a}) should be satisfied for 
any point $r$, and  the boundary conditions (\ref{2-d}) and (\ref{2-e}) 
should be also satisfied at the position of a brane ($r=r_i$).
Substituting Eqs.  (\ref{2-d}) and (\ref{2-e}) into Eq. (\ref{2-a}),
we find that  $k$ must vanish. Therefore,
 we construct only an instanton solution with a flat brane.  
This condition, $k=0$, may be understood from the following fact.
In the present model, 
the effective cosmological constant on the brane
is given by 
\begin{eqnarray}
 ^{(4)}\Lambda = \frac{1}{4} \left[V(\phi)+\frac{1}{12} \lambda(\phi)^2
 -\frac{1}{8}\left(\frac{d\lambda(\phi)}{d\phi} \right)^2 \right]\,, 
 \label{2-g}
\end{eqnarray}
(see \cite{Maeda-Wands00}).
Substituting our potential form (\ref{2-h}) and tension (\ref{2-j})
 into Eq. (\ref{2-g}), we find that
the effective 4-dimensional cosmological constant vanishes.
Therefore the brane should be flat.
This is because of supersymmetry.
In what follows,
we restrict our analysis to a flat-brane model.

\section{FLAT-BRANE INSTANTON} \label{flat-brane}
In this section, we present  flat-brane instantons.
We discuss a two-brane model and a single-brane one separately.

\subsection{Two-brane instanton}
Here we construct a two-brane instanton solution.
We solve the equations of motion in the bulk, i.e.,
Eqs.  (\ref{2-b}) and (\ref{2-c}).
The solutions are given by
\begin{eqnarray}
 b(r) = b_0 r^{\frac{1}{6\beta^2}}, \label{4-a}
\end{eqnarray}
\begin{eqnarray}
 \phi(r) = \frac{1}{\beta} \ln \left(\sqrt{2} v \beta^2 r \right).
 \label{4-b}
\end{eqnarray}
When the tension of the first brane is negative and the second brane is
positive, respectively, 
this bulk solution satisfies the jump conditions (\ref{2-d}) and (\ref{2-e}) 
at any point $r$. 
In other words, the positions of branes are freely chosen.
We could put branes anywhere we prefer.
This is because branes are flat (no curvature).
 
In order to find the most preferable positions
of branes, we evaluate the Euclidean action (\ref{2-k}).
The Euclidean action (\ref{2-k}) is rewritten 
using equations of motion (\ref{2-a}-\ref{2-b}) as
\begin{eqnarray}
 S_E &=& -\frac{1}{2\kappa^2_5} \left\{
 \int_{{\mathcal M}} d^5x\sqrt{g}\left[\frac{2}{3}V(\phi) \right]  \right. 
 \nonumber \\
 & & \left. 
 +\sum_{i}\int_{r=r_i}d^4x \sqrt{g_{(i)}}  \left[\frac{1}{3}
  \lambda_{(i)}(\phi) \right] \right\} \nonumber 
\end{eqnarray}
\begin{eqnarray}
 &=&  -\frac{V^{\gamma}_{4}}{2\kappa^2_5} \left\{
  2 \int^{r_2}_{r_1} dr b^4(r) \frac{2}{3}V(\phi) \right.  \nonumber \\
 & &\left. +\sum_{i} \frac{1}{3} b^4(r) \lambda_{(i)}(\phi) \right\}, \label{4-c}
\end{eqnarray}
where $V^{\gamma}_{4}=\int dx^4 \sqrt{\gamma}$ is the volume of 
the manifold with a 4-metric 
$\gamma_{\mu \nu}$. The factor 2 in front of the integral in
(\ref{4-c}) is required since our instanton solution is 
constructed by two copies of
Euclidean manifolds (see Fig. \ref{fig1}).
Performing the integration with solutions (\ref{4-a}) and (\ref{4-b}),
we obtain
\begin{eqnarray}
 S_E &=& - \frac{V^{\gamma}_4}{2\kappa^2_5} \frac{b_0^4}{\beta^2}
 \left\{\frac{2}{3}\left( r_1^{\frac{2}{3\beta^2}-1}
 - r_2^{\frac{2}{3\beta^2}-1} \right) \right. \nonumber \\
  & &\left. - \frac{2}{3} \left(r_1^{\frac{2}{3\beta^2}-1}
 -r_2^{\frac{2}{3\beta^2}-1} \right)
 \right\} = 0.
\end{eqnarray}
Unfortunately, this action vanishes and
then does not have any minimum (or maximum)
with respect to the positions of branes 
($r_1$ and $r_2$).
Hence the positions of branes are not determined 
by the least action principle.
This result may be related to the 
problem of moduli stabilization.
We may need some additional
mechanism to stabilize 
the distance between two branes \cite{moduli_stabilization}.

\subsection{Singular single-brane instanton}

In this section, we consider the singular instanton such as the 
Hawking-Turok instanton \cite{Hawking-Turok98}. 
For a single-brane instanton,
from the bulk solution (\ref{4-a}), 
we find that $b$ vanishes at $r=0$.
At this point, $r=0$, the spacetime curvature of the 5-dimensional
manifold diverges and a scalar field also dose so. 
Therefore this instanton solution has a singularity at $r=0$.
However, this singularity is ``mild'' since the Euclidean action of this
instanton solution is finite for some range of $\beta$.

To show it explicitly, we evaluate the Euclidean action.
First we divide the action into three terms:
$S_E^{\rm total}=S_E^{\rm bulk}+S_E^{\rm brane}+S_E^{\rm singularity}$,
where
$S_E^{\rm brane}$ also includes contribution from the 
Gibbons-Hawking term \cite{Gibbons-Hawking77} at the positions of branes.
$S_E^{\rm singularity}$ is a boundary term at the singularities. 
Although we do not know an appropriate boundary condition at a
singularity, 
our situation is the same as the case of a 4D singular instanton.
For the 4D singular instanton, Vilenkin \cite{Vilenkin:1998pp} adopted the
Gibbons-Hawking term  at the singularity. 
Here we also adopt the Gibbons-Hawking term.
We will discuss later on the ambiguity of boundary condition at
a singularity.

\begin{figure}[t]
\begin{center}
 \includegraphics[height=4cm]{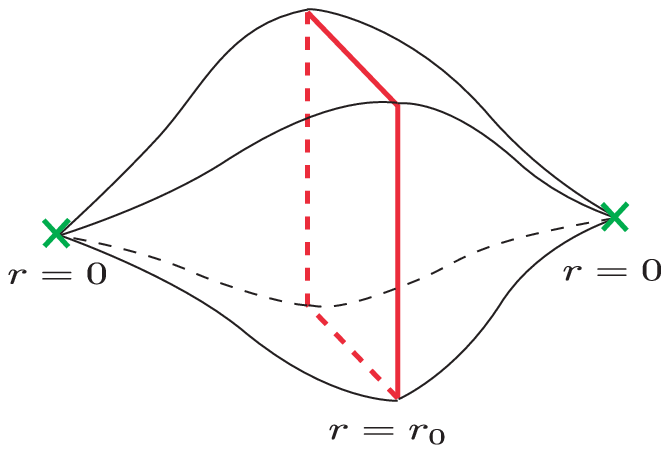}
 \caption[fig2]{Singular brane instanton. 
The thick vertical line at $r=r_0$ represents the $T^4$ brane at which
the two identical 5-dimensional bulk spaces are glued. The cross marks at $r=0$
 represent singular points. } \label{fig2}
\end{center}
\end{figure}

In order that we perform the integration for $[0,r_0]$,
first we integrate for an interval $[\varepsilon,r_0]$ ($\varepsilon
\ll r_0$) and we then take a limit of
$\varepsilon \rightarrow 0$.
The Euclidean action of the bulk is given by
\begin{eqnarray}
 S_E^{\rm bulk} &=& -\frac{1}{2\kappa_5^2}\int_{\mathcal M} dx^5 \sqrt{g}
  \left[R -\frac{1}{2}\partial_a \phi \partial^a \phi - V(\phi) \right]
  \nonumber  \\
  &=& -2\frac{V_4^{\gamma}}{2\kappa_5^2} \lim_{\varepsilon\rightarrow 0}
   \int_{\varepsilon}^{r_0}\frac{2}{3} b(r)^4 V(\phi)dr \label{4-d}  \\
 &=& \lim_{\varepsilon\rightarrow 0}\frac{V_4^{\gamma }}{2\kappa_5^2}
  \frac{4b_0^4}{6\beta^2} 
  \left(r_0^{\frac{2}{3\beta^2}-1}-\varepsilon^{\frac{2}{3\beta^2}-1}
  \right),
\end{eqnarray}
where the factor 2 in (\ref{4-d}) is required since our instanton 
contains two copies of
the Euclidean manifold (see Fig. \ref{fig2}).

The brane action is evaluated as
\begin{eqnarray}
 S_E^{\rm brane} &=& -\frac{1}{2\kappa_5^2}\int_{r=r_0}dx^4
 \sqrt{g(i)}\left[K^{\pm}_{r=r_0}-\lambda_0(\phi) \right] \nonumber  
 \\
 &=&  -\frac{V_4^{\gamma}}{2\kappa_5^2} \frac{1}{3}b(r_0)^4
 \lambda_0(\phi) \nonumber \\ 
 &=& -\frac{V_4^{\gamma}}{2\kappa_5^2}  \frac{2}{3\beta^2}b_0^4
 r_0^{\frac{2}{3\beta^2}-1}. 
\end{eqnarray} 
Finally 

The Gibbons-Hawking term at the singularity is given by
\begin{eqnarray}
 S_E^{\rm singularity}&=& -2 \lim_{\varepsilon \rightarrow
  0}\int_{r=\varepsilon} dx^4 
  \sqrt{g_{(4)}} \left[\frac{1}{\kappa_5^2} K \right] \label{4-e} \\
 &=& \lim_{\varepsilon \rightarrow
  0}\frac{V_4^{\gamma}}{\kappa_5^2}\frac{4b_0^4}{3\beta^2}
  \varepsilon^{\frac{2}{3\beta^2}-1}, 
\end{eqnarray}
where the factor 2 in (\ref{4-e}) is also required since this instanton
 solution has two singularities.

We find that the total action is evaluated as
\begin{eqnarray}
 S_E^{\rm total}=\frac{V_4^{\gamma}}{\kappa_5^2}\frac{b_0^4}{\beta^2}
 \varepsilon^{\frac{2}{3\beta^2}-1} \Big|_{\varepsilon \rightarrow0}.
\end{eqnarray} 
If $\beta^2 \leq 2/3$, this action converges.
In that case, our instanton with a singularity could be 
allowed, which is called a singular instanton.

This action also has no minimum with respect to $r$.
Therefore the position
of a brane is not fixed, just as 
the case of a two-brane instanton.

Although we have adopted the Gibbons-Hawking term at the singularity,
the boundary condition at the singularity is somewhat
 ambiguous.
For instance,
when we consider the action of the asymptorically flat or asymptotically
AdS spacetime,
we need to add the counter term in order that the action is well-difined 
\cite{Hawking:1995fd,Kraus:1999di,Mann:2005yr}.

However, in our present model, even 
if the Gibbons-Hawking term at the singularity is dropped from the
action, our result does  not be changed significantly.
The action converges for $\beta^2 \leq 2/3$,
which is the same condition as the case with the Gibbons-Hawking term. 
Moreover, when we calculate the quadratic term of the perturbed action
(fluctuations around the solution),
which will be explicitly shown 
in next section, it will converges.
This guarantees that such an instanton solution is well-defined
and harmless,
just as the Hawking-Turok instanton is.
Therefore we may conclude that our result does not depend
on the choice of the boundary condition.

\section{Quadratic term of the perturbed action} \label{2nd-action}
Here, 
we investigate perturbations around a
singular instanton solution,
whose Euclidean action is finite.
Although a singular instanton 
contains a singularity,
it is well-defined if
fluctuations around the instanton solution
do not diverge near singularity.
Such an instanton solution could be 
realized under some circumstance \cite{Lavrelashivili98}.

We then calculate the quadratic term of the perturbed action
 near the singular point.
As a useful form of a singular instanton
 to calculate the variations, we adopt the conformal frame, i.e.,
\begin{eqnarray}
 ds^2_E=b^2(R) \left[dR^2+\gamma_{\mu \nu}dx^{\mu}dx^{\nu}
	       \right]. \label{metric}
\end{eqnarray}
Under this coordinate system, the singular instanton solution
(Eqs. (\ref{4-a}) and (\ref{4-b})) is given by: \\
\begin{eqnarray}
 {\rm for} ~~\beta^2&\neq& 1/6,
\nonumber\\
b(R)&=& \tilde{b}_0 \left[\left(1-\frac{1}{6\beta^2}\right)
R
	       \right]^{\frac{1}{6\beta^2-1}}\,, \\
 \phi(R)&=&\phi^c_1+\frac{6\beta^2}{6\beta^2-1}  
  \ln \left[\left(1-\frac{1}{6\beta^2}\right)R\right]\,, 
\\[1em]
{\rm for} ~~\beta^2&=& 1/6, 
\nonumber
\\[.5em]
b(R) &=& b_0 e^{b_0 R}, \\
 \phi(R) &=& \phi^c_2+\frac{1}{\beta} R,
\end{eqnarray}
where $\tilde{b}_0$, $\phi^c_1$, and $\phi^c_2$ are constants.
For simplicity, we have introduced new variable 
$\phi(R)\equiv \sqrt{2}\kappa_5\varphi(R)$.   
Note that the singular point of the instanton 
corresponds to $R = 0$ for $\beta^2>1/6$, while  
$R = -\infty$ for $\beta^2\leq 1/6$.

The detailed calculation is given in Appendix A.
Here we show just the result.
Up to total divergence terms, 
the quadratic term of the perturbed action (\ref{2nd-vari}) is
\begin{eqnarray}
  \delta_2 S_E = \frac{1}{2} \int d^5x \left[\dot{f}^2+f_{|\mu}f^{|\mu}
	        + \frac{\ddot{z}}{z} f^2 \right], \label{2nd-vari2}
\end{eqnarray}
where a dot ($\dot{~}$) denotes the derivative with respect to $R$,
a vertical line ($|$)denotes the covariant derivative with respect to 
$\gamma_{\mu\nu}$,
$
f\equiv
b^{\frac{3}{2}}\left[\delta\varphi+(\dot{\varphi_0}/\mathcal{H})\psi \right]
$
 is a gauge-invariant combination of
perturbations of a scalar field and of the metric, and 
$z\equiv
b^{3/2}\dot{\varphi}_0/\mathcal{H}$.
$\mathcal{H}=\dot{b}/b$ and $\psi$ are the Hubble parameter
and one of the scalar modes of metric perturbations 
(see Eq. (\ref{paramet})), respectively.

For our case,
\begin{eqnarray}
 \frac{\ddot{z}}{z} &=& \frac{3}{4}\frac{5-12\beta^2}{(6\beta^2-1)^2 R^2}
 \ \ \ \mbox{for} \ \ \beta^2 \neq \frac{1}{6} \\ 
  &=& \frac{9}{4}b_0^2  ~~~~~~~~~~~~~~~~~~  \mbox{for} \ \ \beta^2=\frac{1}{6} .
\end{eqnarray}
Eq. (\ref{2nd-vari2}) is exactly the same as
the form of a scalar field in flat spacetime
with a $r$-dependent mass $\ddot{z}/z$.
Note that $\gamma_{\mu\nu}$ is the metric of 4-dimensional
flat space.
For $\beta^2>5/12$, the ``mass" term is negative. 
Hence we expect that such an  instanton is unstable
with respect to ``time" $r$.
However, there might be an instanton solution
with different metric ansatz (e.g., with 
different topology), whose action is
 lower that of the present solution.
Although we may be able to 
conclude that there is no instanton for $\beta^2>5/12$,
it is unlikely.
Hence, in what follows,
 we consider only the case of $\beta^2\leq 5/12$.

Varying Eq. (\ref{2nd-vari2}) with respect to $f$,
we obtain  the equation of
motion for $f$: 
\begin{eqnarray}
\ddot{f}+\Delta f -(\ddot{z}/z) f=0\,.
\end{eqnarray}
The separation of variable, i.e., 
$f=f_\ell(R)Y_\ell(x^{\mu})$, 
gives two ordinary differential equations:
\begin{eqnarray}
  (\Delta + \ell^2)Y_\ell=0\,, \\
 \ddot{f}_\ell-\left(\ell^2+\frac{\ddot{z}}{z} \right)f_\ell=0\,,
\label{eq-vk}
\end{eqnarray}
where $\ell$ is an eigen value and $Y_\ell$ is its
eigen function.

For each eigen mode $\ell$ , the quadratic term of the perturbed action is
given by
\begin{eqnarray}
 \delta_2 S_E=\frac{1}{2}\int dR
  \left[\dot{f}_\ell^2+\ell^2+\frac{\ddot{z}}{z} \right]\int d^4x
~Y_\ell^2.
\label{SE2_singular}
\end{eqnarray}
Analyzing the behavior of $f_\ell$ near the singularity,
we evaluate this action.
For $\beta^2>1/6$, toward the singularity ($R\rightarrow 0$),
the $\ell^2$ term dominates in Eq. (\ref{eq-vk}).
Then the regular solution of $f_\ell$ behaves as 
\begin{eqnarray}
 f_\ell \propto R^{\frac{3}{2(6\beta^2-1)}}. 
\label{vk1}
\end{eqnarray}
Inserting this solution into the action (\ref{SE2_singular}),
we obtain
\begin{eqnarray}
 \delta_2 S_E \propto
  R \ell^2+\frac{3C_1}{2(6\beta^2-1)}R^{\frac{2(2-3\beta^2)}{6\beta^2-1}}, \label{Svk1}
\end{eqnarray}
where $C_1$ is an integration 
constant. 
Near the singularity ($R\rightarrow0)$, (\ref{Svk1}) is finite
if  $1/6<\beta^2\leq2/3$. 

For $\beta^2<1/6$, near singularity ($R \rightarrow - \infty$),
the $\ddot{z}/z$ term dominates in Eq. (\ref{eq-vk}).
The regular solution of $f_\ell$ behaves as
\begin{eqnarray}
 f_\ell \propto e^{\omega R}, 
\end{eqnarray} 
where $\omega^2=\ell^2$.
We also find the same form of the solution  
for $\beta^2=1/6$ if we set  $\omega^2=\ell^2+9b_0^2/4$. 

Near the singularity ($R\rightarrow -\infty$),
the integration 
gives the asymptotic behavior of the action as
\begin{eqnarray}
 \delta_2S_E \propto 2\omega^2 e^{2\omega R}. \label{Svk2}
\end{eqnarray}
This is finite near singularity.

We conclude that the singular
instanton is well-defined for $\beta^2 \leq 5/12$. 

\section{Toward an inflating-brane instanton} \label{Pb-dS}

As we have shown in Sec. \ref{action-eqs}, 
 de Sitter brane instanton is not possible 
for the present potential form and  tension 
(Eqs. 
(\ref{2-h}) and (\ref{2-j})).
This is because our potential is based on supersymmetry.
However, de Sitter brane may be preferred 
from the point of view of cosmology.
 Supersymmetry is also broken in the present universe. 
Hence, in this section, 
we examine other potentials which 
contain some correction terms.
Such corrections may be expected from quantum effects
via SUSY breaking process. 
They may provide us an inflating-brane instanton
\cite{dS_instanton}.

To be concrete, we consider the model with 
the following tension:
\begin{eqnarray}
 \lambda=\pm (2\sqrt{2}v e^{-\beta \phi}+\lambda_i), \label{lambda+1}
\end{eqnarray}
where $\lambda_i$ is regarded as a vacuum energy on the $i$-th brane,
e.g., the vacuum energy of quantum 
matter fields on the brane.
We assume 
$\lambda_i$ is independent of a scalar field $\phi$. 
Another model we discuss is the one with the tension  such that 
\begin{eqnarray}
 \lambda= \pm 2\sqrt{2} \alpha v  e^{-\beta \phi}, \label{lambda+2}
\end{eqnarray}
where $\alpha$ describes a deviation from  
the tension of the original model (\ref{2-j}).
This model is equivalent to the model with a modified
bulk potential.
In fact, when we perform  the transformation such that
$v\rightarrow v/\alpha $ and 
$\alpha\rightarrow [(\beta^2-2/3)/(\beta^2-2/3+\delta)]^{1/2}$,
we find that 
the model (\ref{lambda+2}) is the same as the model
with  the bulk potential 
$V=(\beta^2-2/3+\delta)\exp[-2\beta\phi]$, where
 $\delta$ is a deviation from the original 
bulk potential \cite{Koyama-etal}.
 
This two modifications give non-vanishing cosmological 
constant on the brane.
For some range of parameters $\lambda_i$ and $\alpha$
(e.g., $\lambda_i>0$ or $\alpha<1$),
we find a positive cosmological constant, which guarantees
existence of de Sitter solution on the brane.

First we discuss the case of the tension (\ref{lambda+1}).
The junction condition in this case is 
\begin{eqnarray}
 b'(r_i) &=& \frac{b}{12}\left(2\sqrt{2}v 
     e^{-\beta \phi}+\lambda_i \right)  \\
 \phi'(r_i) &=& \sqrt{2}v \beta e^{-\beta \phi}. \label{junc-phi}
\end{eqnarray}
When we define the new variable 
\begin{eqnarray}
 J\equiv \phi'-\sqrt{2}v \beta e^{-\beta \phi}\,,
\end{eqnarray}
the junction condition for a scalar field 
is given by 
$J=0$ at $r=r_i$. 
Using the equations of motion (\ref{2-a})-(\ref{2-c}) and 
the junction condition (\ref{junc-phi}),
the derivative of $J$ with respect to $r$ 
at $r=r_i$ is given by
\footnote{Here we have assumed $b'>0$. 
If the brane at $r=r_1$ ($r=r_2$ or $r=r_0$) 
has negative (positive) tension, 
which is natural ansatz,
we can prove analytically that this condition is 
obtained for $\beta^2>2/3$ .
For $\beta^2<2/3$, we have confirmed it numerically.}
\begin{eqnarray}
 J'(r_i) &=& -\frac{4}{3}\beta v^2 e^{-2\beta \phi} 
        \left(1+\frac{18}{b^2 v^2} e^{2\beta \phi}\right)^{1/2}
         \nonumber \\
	& & +\frac{4}{3}\beta v^2 e^{-2\beta \phi}
         \Bigm|_{r=r_i} < 0.
\end{eqnarray}
$J'$ is always negative at the position of brane.
This means that $J=0$ is satisfied on either the brane at $r=r_1$ or
at $r=r_2$. 
Hence, if $J=0$ on one brane, e.g., $r_1$,
then we cannot impose $J=0$ on the other brane ($r_2$).
This means that we cannot construct two-brane instanton in this model.
For a single brane model, we find $J<0$ at $r=0$ from no boundary
boundary condition.
$J=0$ is not satisfied at any position of
the brane.

Next we consider the model with tension (\ref{lambda+2}).
Then the junction condition is given by
\begin{eqnarray}
 b'(r_i)=\frac{\sqrt{2}}{6}\alpha v e^{-\beta \phi} \Bigm|_{r=r_i} \\
 \phi'(r_i)=\sqrt{2}\alpha v e^{-\beta \phi} \Bigm|_{r=r_i}.
\end{eqnarray} 
We rewrite this condition  as follows:
\begin{eqnarray}
 b'(r_i)&=\frac{1}{6\beta} \phi'(r_i)b(r_i) \\
 \phi'(r_i)&=\sqrt{2}\alpha v e^{-\beta \phi} \Bigm|_{r=r_i}.
\end{eqnarray}
Defining a new variable $W$ by
\begin{eqnarray}
 W\equiv b'(r_i)-\frac{1}{6\beta} \phi'(r_i)b(r_i)\,,
\end{eqnarray}  
and using the equations of motion (\ref{2-b}) and (\ref{2-c}),  
we obtain the derivative of $W$ with respect to $r$ as
\begin{eqnarray}
 W'=-3\frac{b'}{b}W+\frac{3}{b} \Bigm|_{r=r_i}.
\end{eqnarray}
Since $W=0$ at the position of the brane ($r=r_i$),
$W'$ is always positive.
We cannot construct a two-brane instanton solution,
because $W=0$ and $W'>0$ are not satisfied at both boundaries ($r=r_1, r_2$)
simultaneously.
For a single-brane model,
$W=1$ at $r=0$ from no boundary boundary condition. 
In this case $W=0$ is not also satisfied because 
$W'>0$ at any position of brane ($W=0$).

As we have shown above, 
we cannot construct a de Sitter (inflating) brane instanton 
for the present types of modification.
We may need different type of correction terms.

\section{CONCLUSION} \label{conclusion}
We have presented instanton solutions in the model with
a bulk scalar field.
For an exponential potential of a bulk scalar field and tension,
which includes Randall-Sundrum model ($\beta=0$) 
and the 5-dimensional effective Ho\u{r}va-Witten theory 
($\beta=1$),
we construct  flat brane instanton solutions;
one is a brane instanton solution with two flat branes,
and the other is  that with a single brane.
We find that a single brane instanton always has a singularity.
However, the Euclidean action of such an instanton
is finite.
As a result,  this singular instanton could be realized
just as the Hawking-Turok singular instanton.
In order to guarantee that  the instanton
is well-defined, 
we also analyze the behaviour of the action 
 perturbed 
around the singular instanton. 
We find   that the quadratic term of the  perturbed action 
 is finite if $\beta^2<2/3$.
 This guarantees  the singular instanton is well-defined.
The second variation equation of the perturbed action is
the same form as that of 
a scalar field in flat spacetime with ``time ($R$)''-dependent mass.
For $\beta^2>5/12$, this mass term becomes negative,
which probably means that the instanton is unstable.
We then conclude that a singular single-brane instanton
is possible if $\beta^2\leq 5/12$.

The action of the instanton solutions
does not have any minimum with respect to the positions of 
branes. Then we cannot adopt the least action principle to
predict the initial state of a brane
universe.

Taking into account some quantum corrections for a bulk potential
or tension via SUSY breaking process,
we have also investigated possibility of de Sitter brane 
instanton solution.
However, we could not find appropriate
instanton solutions.
In order to obtain de Sitter brane instanton,
we may need to include other important effects,
such as the Casimir energy, which are not take into 
account here, or higher curvature correction terms
discussed in \cite{Aoyanagi-Maeda04}. 

For a
flat brane instanton we constructed,
we have to consider the evolution of the brane universe 
after its creation.
We may not need inflation
just after creation \cite{Linde},
or may have the KKLMMT type inflation \cite{KKLMMT}. 
These issues are left to future study.

~\\
\acknowledgments

We would like to thank S. Mizuno for useful discussion.
K. A. acknowledges T. Torii and N. Okuyama for valuable comments.
This work was partially supported by the Grant-in-Aid for Scientific Research
Fund of the JSPS (No. 17540268) and another for the 
Japan-U.K. Research Cooperative Program,
and by the Waseda University Grants for Special Research Projects and 
 for The 21st Century
COE Program (Holistic Research and Education Center for Physics
Self-organization Systems) at Waseda University.

\appendix
\begin{widetext}
~
\vskip .5cm 
\section{Quadratic term of the perturbed action}

We begin with the original Euclidean action:
\begin{eqnarray}
 S_E=-\int d^5x \sqrt{g}
  \left[\frac{1}{2\kappa_5^2} R-\frac{1}{2}\nabla_{a}\varphi \nabla^{a}
   \varphi - V(\varphi)  \right]. 
\end{eqnarray}
We  perturb metric and a scalar field  as 
\begin{eqnarray}
 g_{a b}=g^{(0)}_{a b}+h_{a b}, ~~~~~
  \varphi = \varphi_0+ \delta \varphi.
\end{eqnarray}
We then expand the total action until second order perturbations as
\begin{eqnarray}
 \delta_2 \left(\sqrt{g}R \right)
  =\sqrt{g^{(0)}} \left\{h^{a}_{c} h^{c b} 
   -\frac{1}{2}h h^{a b} 
   +(\frac{1}{8}h^2-\frac{1}{16}h^{c}_{d} h^{d}_{c})
   g^{(0) a b} \right\} R^{(0)}_{a b} \nonumber  \\
  -\sqrt{g^{(0)}}\left(h^{a b}-\frac{1}{2}h g^{(0)a b}
		  \right) \delta R_{a b} 
  +\sqrt{g^{(0)}} g^{(0)a b}\delta^2 R_{a b},
\end{eqnarray}
where 
\begin{eqnarray}
 \delta R_{a b}=\delta \Gamma^{c}_{a b ; c}
  -\delta \Gamma ^{c}_{a c ;b}
  +2\left(\delta \Gamma^{c}_{a d} \delta 
     \Gamma^{d}_{c b} 
    -\delta \Gamma^{d}_{d c} \delta
     \Gamma^{c}_{a b}\right),
\end{eqnarray}
\begin{eqnarray}
\delta^2 R_{a b}=
\delta^2\Gamma^{c}_{a b ;c}-\delta^2 \Gamma^{c}_{a
 c ;b}+\delta \Gamma^{d}_{c d}\delta
 \Gamma^{c}_{a b}-\delta \Gamma^{d}_{c b}
 \delta\Gamma^{c}_{a d},
\end{eqnarray}
and
\begin{eqnarray}
 \delta \Gamma^{c}_{ab}=
  \frac{1}{2} \left\{h^{a}_{\ c ;b}+h_{\ b;a}^{c}
	      -h_{ab}^{\ \ \ c}\right\}.
\end{eqnarray}
\begin{eqnarray}
 \delta^2 \left(\sqrt{g} R \right) &=&
  \sqrt{g^{(0)}} 
  \left[\left(h^{ac}h_{c}^{\
	 b}-\frac{1}{2}hh^{ab}\right)R^{(0)}_{ab} 
  +\left(\frac{1}{8}h^2 
  -\frac{1}{4}h_{cd}
    h^{cd} \right)R^{(0)}
  \right.   \nonumber \\
 & &  \left.  
       + \frac{1}{4}\left(2h^{ab;c}h_{ac;b}
	       -h^{ab;c}h_{ab\;c}
	       -2h^{ab}_{\ \ \ ;b}h_{;a}+h_{;a}h^{;a}
	      \right)\right].
\end{eqnarray}
\begin{eqnarray}
 & &\delta^2 \left(\sqrt{g} \left[-\frac{1}{2}g^{ab}\partial_{a} 
   \varphi_0
   \partial_{b}\varphi_0-V(\varphi_0) \right]\right) 
   \nonumber \\ 
 &=& \sqrt{g^{(0)}}\left[\left(-\frac{1}{2}h^{a}_{\ c}h^{cb}
  +\frac{1}{4}h h^{ab}\right)\partial_{a}\varphi_0 \partial_{b}\varphi_0 
  +\left(\frac{1}{8}h^2-\frac{1}{4}h_{cd}h^{cd}\right)
  \left(-\frac{1}{2}g^{(0) a b} \partial_{a} \varphi_0 \partial_{b}
   \varphi_0 -V(\varphi_0) \right) 
  \right. \nonumber \\ 
 & & \left.  - \frac{1}{2}g^{(0)ab}\partial_{a}\delta\varphi
  \partial_{b}\delta\varphi +\left(h^{ab}
  -\frac{1}{2}h g^{(0)ab}\right) \partial_{a}\varphi_0
  \partial_{b}\delta\varphi -\frac{1}{2}h \frac{dV}{d\varphi}
\Bigm{|}_0\delta\varphi 
  -\frac{1}{2}\frac{d^2 V}{d\varphi^2}\Bigm{|}_0\delta\varphi^2 \right]
\end{eqnarray}
With the metric ansatz (\ref{metric}), 
a background spacetime is given by the equations of motion:
\begin{eqnarray}
&& \mathcal{H}^2-\dot{\mathcal{H}}-k=\frac{\kappa^2_5}{3}\dot{\varphi}_0^2, \\
&& \dot{\mathcal{H}}+\mathcal{H}^2-k=
-\frac{\kappa^2_5}{6}\left[\dot{\varphi}_0^2+2b^2V(\varphi_0) \right], \\
&& \ddot{\varphi}_0+3\mathcal{H}\dot{\varphi}_0-b^2V_{,\varphi}(\varphi_0)=0,
\end{eqnarray}
where a dot denotes the derivative with respect to $R$,
$k$ is the curvature of the brane, and $\mathcal{H}=\dot{b}/b$
is the ``Hubble'' parameter. 
Using these equations, the second order variation is given by 
\begin{eqnarray}
 & &-\frac{1}{2\kappa_5^2}\delta^2\left(\sqrt{g}R \right)
 - \delta^2 \left(\sqrt{g} \left[-\frac{1}{2}g^{ab}\partial_{a} \varphi
   \partial_{b}\varphi-V(\varphi) \right] \right)  
  \nonumber \\
 & &=-\sqrt{g^{(0)}}
  \left[-\frac{1}{8\kappa^2}  
   \left(h^2-2h_{cd}h^{cd}\right)\frac{1}{b^2} 
    \left(\dot{\mathcal{H}}+3\mathcal{H}^2-3k \right)  
  +\frac{1}{8\kappa^2}\left(2h^{ab;c}h_{ac;b}
	       -h^{ab;c}h_{ab;c}
	       -2h^{ab}_{\ \ \ ;b}h_{;a}+h_{;a}h^{;a}
	      \right)  \right. \nonumber \\ & & ~~~~~~~~~~~~~~~~~~ \left.
   -\frac{1}{2}g^{(0)ab}\partial_{a}\delta\varphi
  \partial_{b}\delta\varphi +\left(h^{ab}
  -\frac{1}{2}h g^{(0)ab}\right) \partial_{a}\varphi
  \partial_{b}\delta\varphi -\frac{1}{2}h \frac{dV}{d\varphi}\Bigm{|}_0
\delta\varphi 
  -\frac{1}{2}\frac{d^2 V}{d\varphi^2}\Bigm{|}_0\delta\varphi^2
  \right] 
\end{eqnarray} 

Since scalar, vector and tensor modes of perturbations 
are decoupled in our model,
we focus only on scalar perturbations.
The scalar mode of metric perturbations is described as
\begin{eqnarray}
 ds^2_E=b^2(R) \left[\left(1+2A)dR^2 +2B_{|\mu} dx^{\mu}dR 
        +\left\{(1-2\psi)\gamma_{\mu \nu}+2E_{|\mu \nu} \right\}
		dx^{\mu} dx^{\nu}    \right)\right], \label{paramet}
\end{eqnarray}
where $A$, $B$, $E$, and $\psi$ are metric components of perturbations.

The quadratic term of action for scalar perturbations reads
(see \cite{Mukhanov-etal92,Garriga-etal97})
\begin{eqnarray}
 \delta_2 S_E &=& -\frac{1}{2\kappa^2_5} \int d^5x \sqrt{\gamma}
  \left\{ b^3
  \left[ 12\dot{\psi}^2+24\mathcal{H}A\dot{\psi}+3(3\mathcal{H}^2
   +\dot{\mathcal{H}})A^2 -6\psi_{|\mu}(A-\psi)^{|\mu}
  \right. \right. \nonumber \\ & & \left. \left.
  +2\kappa^2_5 \left(-\dot{\varphi}_0\dot{A}\delta\varphi
		-4\dot{\varphi}_0\delta\varphi\dot{\psi}
		-2A b^2\frac{dV}{d\varphi}\delta\varphi  \right)
  +\kappa^2_5 \left(-\delta\dot{\varphi}^2-\delta\varphi_{|\mu}
\delta\varphi^{|\mu}
	       -\frac{d^2V}{d\varphi^2}\delta\varphi^2 b^2 \right) 
	\right. \right. \nonumber \\ & & \left. \left.
	+6\bigtriangleup(B-\dot{E}) 
	\left(\mathcal{H}A
	 -\frac{\kappa^2_5}{3}\dot{\varphi}_0
	 \delta \varphi+\dot{\psi} \right) 
	+k\left(3A^2+24A\psi-24\psi^2-3(B-\dot{E})\Delta(B-\dot{E}) \right) 
\right]
     \right. \nonumber \\ & & \left.  
     +\mathcal{D}_1+\mathcal{D}_2 \right\}, \label{AforS}
\end{eqnarray}
where $\mathcal{D}_1$ and $\mathcal{D}_2$ are a total divergence terms:
\begin{eqnarray}
 \mathcal{D}_1 &=& {\partial\over \partial R}
                 \left\{b^3 
		\left[- 4\mathcal{H}A^2 -8\mathcal{H}A\psi
		 +2\mathcal{H}A \Delta E +16\mathcal{H}\psi^2
		 -8\mathcal{H}\psi\Delta E
		 +\mathcal{H}(\Delta E)^2-\mathcal{H}B_{|\mu}B^{|\mu} 
		 - 8\mathcal{H}\psi^2 +4\mathcal{H}\psi\Delta E
	        \right. \right.  \nonumber \\ & & \left. \left.
		 -2\mathcal{H}E_{|\mu\nu}E^{|\mu\nu}+A_{|\mu}B^{|\mu}
-4\psi\Delta B
                 +\Delta E \Delta B 
		 +8\kappa_5^2 \psi  \dot{\varphi}_0 \delta \varphi
		 -2\kappa_5^2 \Delta E \dot{\varphi}_0
		 \delta\varphi
		 +2\kappa_5^2 A\dot{\varphi}_0 \delta\varphi 
	\right] \right\},
\end{eqnarray}
and 
\begin{eqnarray}
 \mathcal{D}_2 &=& \left\{b^3
		    \left[\Delta B B^{|\mu}-2\Delta B\dot{E}^{|\mu}
		     +\Delta \dot{E}\dot{E}^{|\mu}+E^{|\mu\nu\rho}E_{|\nu\rho}
		     -E^{|\rho}_{\ \ |\nu\rho}E^{|\mu\nu} -\dot{A}B^{|\mu}
 +B^{|\mu\nu}B_{|\nu}
		     -\Delta BB^{|\mu} 
		    \right. \right.\nonumber \\ & &\left. \left.
		    +4\psi\dot{B}^{|\mu} 		    
		    -\Delta E\dot{B}^{|\mu} +12\mathcal{H}\psi B^{|\mu}
		    -3\mathcal{H}\Delta EB^{|\mu} -6\mathcal{H}AB^{|\mu}
		    +2\kappa_5^2 \delta\varphi \dot{\varphi}_0 B^{|\mu}
		    \right. \right. \nonumber  \\ & &\left. \left.
		    +k\left(12\psi E^{|\mu} -6AE^{|\mu} -E^{|\mu\nu}E_{|\nu}
		       -2\Delta EE^{|\mu}+3(B-\dot{E})(B-\dot{E})^{|\mu}
	\right)
							    \right] \right\}_{|\mu}.
\end{eqnarray}
Here we consider the model with $k=0$ because we obtain only 
a flat brane solution.
By varying Eq. (\ref{AforS}) with respect to $(B-\dot{E})$, we get the
following constraint equation:
\begin{eqnarray}
 \dot{\psi}=\frac{\kappa^2_5}{3}\dot{\varphi}_0 \delta\varphi -\mathcal{H}A \
  . \label{constraint}
\end{eqnarray}
We introduce  a gauge-invariant combination of
perturbations of a scalar field and of the metric \cite{Mukhanov-etal92}
\begin{eqnarray}
 f=b^{\frac{3}{2}}\left[\delta\varphi+(\dot{\varphi_0}/\mathcal{H})\psi \right]
  =b^{\frac{3}{2}}\left[\delta\varphi^{\rm (GI)}
   +(\dot{\varphi_0}/\mathcal{H})\Psi
    \right], \label{potential-v}
\end{eqnarray}
where $\delta\varphi^{\rm (GI)}$ is the gauge-invariant scalar field
perturbation.  
Using (\ref{constraint}) and (\ref{potential-v}) to replace $\dot{\psi}$
and $\delta\varphi$ with $A$, $\psi$ and $v$,
we obtain the following quadratic action:
\begin{eqnarray}
 \delta_2 S_E = \frac{1}{2} \int d^5x \left[\dot{f}^2+f_{|\mu}f^{|\mu}
	        + \frac{\ddot{z}}{z} f^2 +\mathcal{D}_1 +\mathcal{D}_2 
	        +\mathcal{D}_3 \right], \label{2nd-vari}
\end{eqnarray}
where $z=b^{\frac{3}{2}}\dot{\varphi}_0/\mathcal{H}$ and $\mathcal{D}_3$
is a total divergence term:
\begin{eqnarray}
 \mathcal{D}_3 &=& {\partial \over \partial
  R}\left\{2b^{\frac{3}{2}}\dot{\varphi}_0 Af 
		-2b^3\frac{\dot{\varphi}_0^2}{\mathcal{H}}A\psi
		-\frac{3}{2}\mathcal{H}f^2
		-\frac{\kappa_5^2}{3}\frac{\dot{\varphi}_0^2}{\mathcal{H}}f^2
	       \right. \nonumber \\ & & \left.
 		-2b^{\frac{3}{2}}\frac{\ddot{\varphi}_0}{\mathcal{H}}f\psi
		+2b^{\frac{3}{2}}\dot{\varphi}_0 f\psi
		-b^3\frac{\dot{\varphi}_0^2}{\mathcal{H}}\psi^2
		+b^3\frac{\dot{\varphi}_0 
		\ddot{\varphi}_0}{\mathcal{H}^2} \psi^2
		-\frac{3b^2}{\kappa_5^2}\frac{\psi_{|\mu} \psi^{|\mu}}{\mathcal{H}}
	       \right\}.
\end{eqnarray}

This Euclidean action is the same as that of
a massive scalar field $f$ with ``time ($R$)''-dependent mass term, which 
is $({\ddot{z}}/{z})f^2 $, in a 5-dimensional flat Euclidean space.

\end{widetext}



\begin{thebibliography}{00}
\bibitem{dS_instanton}
	 If we include higher curvature terms such as Gauss-Bonnet term,
	 which are also expected to exist as quantum correction terms,
	 we can construct de   Sitter-brane instanton. See \cite{Aoyanagi-Maeda04}
\end{thebibliography}
\end{document}